\documentclass[aps,prb, preprint, citeautoscript, showpacs,floatfix,superscriptaddress]{revtex4-2}
\usepackage{bm}
\usepackage{amssymb}
\usepackage{graphicx}
\usepackage{amsmath}
\usepackage{textcomp}
\usepackage{colordvi}
\usepackage{multirow}
\usepackage{ulem}
\usepackage{braket}
\usepackage{color}
\usepackage{booktabs}
\usepackage{xcolor}

\begin{document}
\title{Gate Voltage Tunable Second Harmonic Generation in Mono- and Bi-layer Black Phosphene}

\author{Kainan Chang} 
\email{knchang@ciomp.ac.cn}
\affiliation{GPL Photonics Laboratory, State Key Laboratory of Luminescence Science and Technology, Changchun Institute of Optics, Fine Mechanics and Physics, Chinese Academy of Sciences, Changchun 130033, China.}
\altaffiliation{These authors contributed equally to this work.}

\author{Yan Meng}
\affiliation{Department of Physics, Institute of Theoretical Physics, University of Science and Technology Beijing, Beijing 100083, China.} 
\altaffiliation{These authors contributed equally to this work.}

\author{Yanyan Qian}
\affiliation{School of Physics, Harbin Institute of Technology, Harbin 150001, China.}
\author{Luxia Wang}
\email{luxiawang@sas.ustb.edu.cn}
\affiliation{Department of Physics, Institute of Theoretical Physics, University of Science and Technology Beijing, Beijing 100083, China.} 
\author{Jin Luo Cheng}
\email{jlcheng@ciomp.ac.cn}
\affiliation{GPL Photonics Laboratory, State Key Laboratory of Luminescence Science and Technology, Changchun Institute of Optics, Fine Mechanics and Physics, Chinese Academy of Sciences, Changchun 130033, China.}

\begin{abstract}
Black phosphorene (BP) has emerged as a promising platform for tunable nonlinear photonics due to its layer-dependent bandgap, high carrier mobility, and remarkable in-plane anisotropy. This study investigates the second-harmonic generation (SHG) of monolayer and bilayer BP under an external static electric field, with describing the electronic states by a tight-binding model and the dynamics by semiconductor Bloch equations. Our results reveal that BP exhibits large second-order nonlinear optical response along the armchair direction, with significant resonant enhancement when the incident photon energy approaches half of its bandgap. Under an applied electric field of $10^7$ V/m, the effective second-order nonlinear susceptibility of BP can be as large as  $10^3$ pm/V, surpassing that of the conventional nonlinear crystal AgGaSe$_2$ by more than an order of magnitude. 
With respect to the static electric field induced by gate voltage,  we discuss the relation between the electric-field-induced second harmonic (EFISH) generation and conventional SHG --
under lower gate voltage, the EFISH approach agrees well with the SHG solutions, whereas the former is no longer applicable under higher gate voltage. 
Specifically, as the increasing gate voltage,
monolayer BP exhibits the bandgap expansion and the corresponding blue-shift in the SHG resonant peak. In contrast, bilayer BP undergoes a semiconductor-to-semimetal transition, forming Dirac cone and generating divergent SHG spectra as photon energy goes to zero. Additionally, the chemical potential allows for precise control over interband and intraband nonlinear responses. This work provides important theoretical foundations for the development of BP-based tunable nonlinear photonic devices and expands the application potential of anisotropic two-dimensional materials in nonlinear optics.
\end{abstract}

\maketitle


\section{Introduction}
\label{introduction}
In recent years, two-dimensional materials have received considerable attention in the field of optics due to their unique band structures and excellent optoelectronic properties.
As an layered semiconductor material, black phosphorene (BP) has become an ideal candidate for advanced optoelectronic systems owing to its strong in-plane anisotropy\,\cite{Yi2019, Wang2016}, layer-dependent bandgap\,\cite{CastellanosGomez2014, Li2017, Chaves2020}, and high carrier mobility\,\cite{Qiao2014a}; and it has shown attractive performance in near- to mid-infrared optoelectronics, with great potential for applications such as photodetectors\,\cite{Chen2017},  field-effect transistors\,\cite{Du2015}, and solar cells\,\cite{Lin2019}. 

Besides the linear optical responses, BP also shows great nonlinear optical responses, such as size-dependent saturable absorption\,\cite{Szydlowska2018, Xu2017}, electro-optic modulators\,\cite{Huang2020}, polarization-sensitive harmonic generation (e.g., third-harmonic generation, THG)\,\cite{Wu2016, Youngblood2017, Hipolito2018}, and field-tunable nonlinear refraction (Kerr effect)\,\cite{Luo2022}, and has become an important nonlinear functional material in integrated photonic devices with applications in ultrafast laser pulse generation\,\cite{Yu2021}, nonlinear frequency conversion\,\cite{Margulis2017}, all-optical ultrafast switching\,\cite{Shi2022}, and optical power limiting\,\cite{Zhang2022}. 
However, all these applications are based on the third-order optical nonlinearities, because BP possesses the inversion symmetry and the electric dipole induced second-order optical nonlinearities are forbidden\,\cite{Boyd2008}, and this limits its applications\,\cite{Sirleto2023} in nonlinear optical devices.
Various methods have been proposed to induce second-order optical nonlinearities in centrosymmetric materials by breaking their inversion symmetry\,\cite{Cheng2017}, including (1) utilizing electric quadrupole-like and magnetic dipole-like effects from spatial electromagnetic field gradients, (2) employing an asymmetric interface or substrate to induce structural symmetry breaking, (3) introducing localized symmetry reduction via strain or curvature, and (4) applying a dc electric field to break inversion symmetry and generate a second-order response via the third-order optical nonlinearity.
Once second-order optical nonlinearity is activated in black phosphorus, it could become an ideal platform for SHG-based applications, such as functionalized BP nanosheet SHG probes for intracellular imaging\,\cite{Peng2021} and oxidation-modulated SHG emission devices\,\cite{Huang2023}.

In this work, our focus is on the electric field induced second harmonic generation (EFISH), which converts two incident photons of frequency $\omega$ to a photon of doubled frequency $2\omega$ with the assistance of applying an external static electric field\,\cite{Ahmed2021}. 
The static electric field $E^c$ breaks the inversion symmetry, and induces second-order optical nonlinearity\,\cite{Jiang2016, Klein2017, Margulis2016}, for which the mechanisms can be classified into two widely discussed types: 
(1) When a weak static electric field is applied, the SHG is usually described by a third-order nonlinear responses involving the static electric field and two optical fields; the effective second-order susceptibility is constructed as $\chi^{(2);dab}(\omega,\omega) = 3\chi^{(3);dabc}(\omega,\omega,0)E^c$ with the superscript indices ``$dabc$'' denoting Cartesian directions $x$, $y$, and $z$\,\cite{Cheng2014, Margulis2016}.
In this case, the static electric field affects not only the band structure, but also the dynamics of the optically excited that cannot be simply attributed to the band structure; this process is originally called as electric-field-induced SHG (EFISH). 
When the material is conductive along the static field direction, the static field leads to a direct current, which can contribute to SHG signal through the current-induced SHG (CISHG). 
(2) When the static electric field is applied along the confined direction (like the $z$ direction for BP), all effects of such field are described in the contents of the band structure\,\cite{Sipe2000}. Thus the static electric field leads to an inversion asymmetric band structure, and the susceptibility of the SHG can be calculated by the same procedure as that of the inversion asymmetric crystal structures. In principle, when the static eletric field is weak enough, the two mechanisms are equivalent.

Furthermore, this work aims to achieve tunable SHG  through dynamic synergistic tuning of electric field strength and chemical potential. Our strategy is based on the following results:
(1)\,Layer number-dependent electric field effects: 
Regarding the band structure, the monolayer BP exhibits electric-field-induced bandgap widening\,\cite{Yang2016, Le2019, Yarmohammadi2021, Le2019b, Pham2019}, while the few-layer BP ($\geq 2$ layers) undergoes band inversion accompanied by the appearance of a Dirac cone\,\cite{Yuan2016, Liu2015, Soleimanikahnoj2017, Dolui2015, Luo2022, Le2018}.
Motivated by these, we investigate the influence of the field on monolayer and bilayer BP to reveal their commonalities and differences in nonlinear optical properties.
(2)\,Chemical potential tuned nonlinearity in doped systems: In graphene, the chemical potential has been shown to maximize the power of EFISH adjusting the doping level\,\cite{Margulis2016}. 
Similarly, in BP, the chemical potential has a huge impact on the direction and magnitude of the anisotropy of linear conductance\,\cite{Xiao2021}.
It is thus anticipated that tuning the doping level will significantly affect the nonlinear susceptibility in EFISH or SHG processes

This paper is organized as follows. In Sec.\,\ref{models}, we introduce the tight-binding model for mono- and bilayer BP, derive the expressions for SHG and EFISH conductivity by the semiconductor Bloch equation, and perform the symmetry analysis on the conductivity tensors. Sec.\,\ref{mono} and Sec.\,\ref{bl} present the EFISH and SHG spectra for monolayer and bilayer BP, respectively, and discuss the regulatory effects of the gate voltage and the chemical potential. We conclude in Sec.\,\ref{conclusions}.

\section{Models}
\label{models}

\subsection{Tight-Binding Method}

BP is a single-element layered crystal composed of phosphorus atoms arranged in a puckered orthorhombic lattice structure. 
Both monolayer and bilayer BP share the same primitive basis vectors $\bm{a} = a{\hat{\bm x}}$, $\bm{b} = b{\hat{\bm y}}$ and $\bm{c} = c{\hat{\bm z}}$, with lattice constants $a = 4.43\,\text{\AA}$ and $b = 3.27\,\text{\AA}$, and $c = 5.46\,\text{\AA}$ \cite{Zhang2023, TaghizadehSisakht2015}.
There are four (eight) atoms in the unit cell of the monolayer (bilayer) BP,
and their atomic positions $\bm{\tau}_\alpha$ are given as follows
$(- \frac{1}{2} + u) \bm a + \frac{1}{4} \bm b + (l - v) \bm c$,
$-  u \bm a + \frac{1}{4} \bm b + (l + v) \bm c$,
$u \bm a - \frac{1}{4} \bm b + (l + v) \bm c$,
$(\frac{1}{2} - u) \bm a - \frac{1}{4} \bm b + (l - v) \bm c$,
with the layer index $l=0$ for the monolayer and $l=0,1$ for the bilayer, and the  parameters  $u\approx0.16$, $v\approx0.20$  \cite{CastellanosGomez2014,Yuan2016, Zare2018, Le2018, Le2019a, Le2019b, Rudenko2015}.

The low energy electronic excitation of pristine BP is described by a tight-binding model, whose parameters are fitted from the first-principle calculation with GW corrections\,\cite{Rudenko2014, Yuan2015, Takao1981}.
The Hamiltonian of monolayer BP\,\cite{Rudenko2014, TaghizadehSisakht2015,Le2019, Pham2019, Hap2024, Le2019c, Yang2016, Zhang2023, Ezawa2014, Yarmohammadi2021} is described by a four-band model
\begin{align}  
	H_{\bm{k}}^{\text{ML}} =H_{\bm k}^0
	+\frac{V}{2}S\,,
\end{align}
with
\begin{align}
H_{\bm k}^0=
\begin{pmatrix}
		0 & A_{\bm{k}} & B_{\bm{k}} & C_{\bm{k}} \\
		A_{\bm{k}}^{*} & 0 & D_{\bm{k}} & B_{\bm{k}} \\
		B_{\bm{k}}^{*} & D_{\bm{k}}^{*} & 0 & A_{\bm{k}} \\
		C_{\bm{k}}^{*} & B_{\bm{k}}^{*} & A_{\bm{k}}^{*} & 0
	\end{pmatrix}\,,\quad
S=\begin{pmatrix}
	-1 & 0 & 0 & 0 \\
	0 & 1 & 0 & 0 \\
	0 & 0 & 1 & 0 \\
	0 & 0 & 0 & -1
	\end{pmatrix}\,,
\end{align}
The matrix elements are given by 
\begin{subequations}
	\begin{align}
		A_{k} &= t_{2}^{\parallel} + t_{5}^{\parallel}e^{-i\bm{k}\cdot\bm{a}}\,, \\
		B_{k} &= 4t_{4}^{\parallel}e^{-i(\bm{k}\cdot\bm{a} - \bm{k}\cdot\bm{b})/2}\cos(\bm{k}\cdot\bm{a}/2)\cos(\bm{k}\cdot\bm{b}/2)\,, \\
		C_{k} &= 2e^{i\bm{k}\cdot\bm{b}/2}\cos(\bm{k}\cdot\bm{b}/2)(t_{1}^{\parallel}e^{-i\bm{k}\cdot\bm{a}} + t_{3}^{\parallel})\,, \\
		D_{k} &= 2e^{i\bm{k}\cdot\bm{b}/2}\cos(\bm{k}\cdot\bm{b}/2)(t_{1}^{\parallel} + t_{3}^{\parallel}e^{-i\bm{k}\cdot\bm{a}})\,.
	\end{align}
\end{subequations}
The Hamiltonian of the bilayer BP is described by an eight-band model\,\cite{Pereira2015}, which can be expressed in block matrix form as:
\begin{align}
H_{\bm{k}}^{\text{BL}} = \begin{pmatrix} H_{\bm k}^0 +Vv S - \frac{V}{2}I& H_{12} \\ H_{12}^\dagger & H_{\bm k}^0 +Vv S + \frac{V}{2}I\end{pmatrix}\,,
\end{align}
with the off-diagonal block 
\begin{align}
	H_{12} &=
	\begin{pmatrix}
		0 & 0 & 0 & 0 \\
		E_{\bm{k}} & 0 & 0 & F_{\bm{k}} \\
		I_{\bm{k}} & 0 & 0 & G_{\bm{k}} \\
		0 & 0 & 0 & 0
	\end{pmatrix}\,,
\end{align} 
accounting for the interlayer coupling, and the the unit matrix $I$.
The interlayer coupling parameters appearing in the off-diagonal blocks are defined as follows:
\begin{subequations}
	\begin{align}
		E_{k} &= t_{1}^{\perp}(1 + e^{i\bm{k} \cdot \bm{b}}) + t_{4}^{\perp}(e^{i\bm{k} \cdot \bm{a}} + e^{i(\bm{k} \cdot \bm{a} + \bm{k} \cdot \bm{b})})\,, \\
		F_{k} &= t_{2}^{\perp}(e^{i\bm{k} \cdot \bm{b}} + e^{i(\bm{k} \cdot \bm{b} - \bm{k} \cdot \bm{a})}) + t_{3}^{\perp}(1 + e^{-i\bm{k} \cdot \bm{a}} + e^{2i\bm{k} \cdot \bm{b}} + e^{i(\bm{2k} \cdot \bm{b} - \bm{k} \cdot \bm{a})})\,, \\
		G_{k} &= t_{1}^{\perp}(1 + e^{i\bm{k} \cdot \bm{b}}) + t_{4}^{\perp}(e^{-i\bm{k} \cdot \bm{a}} + e^{i(\bm{k} \cdot \bm{b} - \bm{k} \cdot \bm{a})})\,, \\
		I_{k} &= t_{2}^{\perp}(1 + e^{i\bm{k} \cdot \bm{a}}) + t_{3}^{\perp}(e^{-i\bm{k} \cdot \bm{b}} + e^{i(\bm{k} \cdot \bm{a} - \bm{k} \cdot \bm{b})} + e^{i\bm{k} \cdot \bm{b}} + e^{i(\bm{k} \cdot \bm{a} + \bm{k} \cdot \bm{b})}).
	\end{align}
\end{subequations}
Here the intralayer hopping energies are $t_{1}^{\parallel} = -1.220$ eV, $t_{2}^{\parallel} = 3.665$ eV, $t_{3}^{\parallel} = -0.205$ eV, $t_{4}^{\parallel} = -0.105$ eV, $t_{5}^{\parallel} = -0.055$ eV and the interlayer hopping energies are $t_{1}^{\perp} = 0.295$ eV, $t_{2}^{\perp} = 0.273$ eV, $t_{3}^{\perp} = -0.151$ eV, and $t_{4}^{\perp} = -0.091$ eV\,\cite{Rudenko2014}.
The gate voltage between different layers is
\begin{align}
V = -eE_{\text{dc}}^cd\,,
\label{VE}
\end{align}
with the electron charge $e=-\left|e\right|<0$, and the $z$-direction displacement of charges $d$ ($d=2vc=2.184$ {\AA} for the monolayer and $d=c=5.46$ {\AA} for the bilayer) \cite{Rudenko2015}.
The eigenstates $C_{n\bm{k}}$ and eigenenergies $\epsilon_{n\bm{k}}$ for band $n$ and wave vector $\bm k$ are obtained by diagonalizing the Hamiltonian through
\begin{align}
\hat{H}_{\bm{k}}C_{n\bm{k}} = \epsilon_{n\bm{k}}C_{n\bm{k}}.
\end{align}
The calculation of the optical properties of BP requires the matrix elements of the position operator $\hat{\bm{r}}_{\bm{k}}$  and velocity operator $\hat{\bm{v}}_{\bm{k}}$ in the eigenstates.
The element of position matrix element in the Bloch basis is written as
\begin{align}	
	{\bm{r}}_{\alpha_1 \alpha_2\bm{k}} &= i \nabla_{\bm{k}} \delta_{\alpha_1 \alpha_2}+ \bm{\tau}_{\alpha_1} \delta_{\alpha_1 \alpha_2}\,,
\end{align}
where $\bm{\tau}_{\alpha_1}$ is the position of atom $\alpha_1$ in the unit cell. 
The  velocity operator is
\begin{align}	
\hat{\bm{v}}_{\bm{k}} = \frac{1}{i\hbar} {[\bm{\hat{\bm{r}}_{\bm{k}}}, \hat{H_{\bm{k}}}]}\,.
\end{align}
The matrix elements of the velocity operator between different energy eigenstates are given by
\begin{align}	
\bm{v}_{n_1n_2\bm{k}} = C^\dagger_{n_1\bm{k}} \hat{\bm{v}}_{\bm{k}} C_{n_2\bm{k}}\,.
\label{v}
\end{align}
Additionally, we define the matrix elements of the position operator as the Berry connection
\begin{align}
	\bm{\xi}_{n_1n_2\bm{k}} = C^\dagger_{n_1\bm{k}} \hat{\bm{r}}_{\bm{k}} C_{n_2\bm{k}}.
    \label{berry}
\end{align}
Since $\hat{\bm{r}}_{\bm{k}}$ contains derivative with respect to $\bm{k}$, the direct calculation for $\bm{\xi}_{n_1n_2\bm{k}}$ from Eq.\,(\ref{berry}) requires that $C_{\bm{k}}$ is a smooth function of $\bm{k}$. However, due to the arbitrariness of phase of the wave function, this is difficult in numerical computations. Generally, the off-diagonal elements of the Berry connection\,\cite{Xiao2010, Sipe2000} can be derived from matrix elements of the velocity operator\,\cite{Aversa1995, Hipolito2016}.
\begin{align}
	\bm{r}_{n_1n_2\bm{k}} = 
	\begin{cases} 
		\bm{\xi}_{n_1n_2\bm{k}} = \frac{\bm{v}_{n_1n_2\bm{k}}}{i\omega_{n_1n_2\bm{k}}} & \text{if } n_1 \neq n_2 \\
		0 & \text{if } n_1 = n_2
	\end{cases}\,,
	\label{r}
\end{align}
with $\hbar\omega_{n_1n_2\bm{k}} = \epsilon_{n_1\bm{k}} - \epsilon_{n_2\bm{k}}$. The diagonal terms $\xi^{a}_{n_1n_1\bm{k}}$ usually appear in the covariant derivative of $(r^{c}_{\bm{k}})_{;n_1n_2k^{a}} = \frac{\partial r^{c}_{n_1n_2\bm{k}}}{\partial k^{a}} - i(\xi^{a}_{n_1n_1\bm{k}} - \xi^{a}_{n_2n_2\bm{k}})r^{c}_{n_1n_2\bm{k}}$, which is alternatively calculated as
\begin{equation}
	(r^{c}_{\bm{k}})_{;n_1n_2k^{a}} = \frac{-ir^{c}_{n_1n_2\bm{k}}{\Delta}^{a}_{n_2n_1\bm{k}} + \hbar {M}^{ca}_{n_1n_2\bm{k}} + i[r^{a}_{\bm{k}}, v^{c}_{\bm{k}}]_{n_1n_2}}{i\omega_{n_1n_2\bm{k}}}\,,
\end{equation}
with ${\Delta}^{a}_{n_2n_1\bm{k}} = v^{a}_{n_2n_2\bm{k}} - v^{a}_{n_1n_1\bm{k}} = \frac{\partial\omega_{n_2n_1\bm{k}}}{\partial k^{a}}$ and
\begin{equation}
	{M}^{ca}_{n_1n_2\bm{k}} = {C}^{\dagger}_{n_1\bm{k}}\frac{1}{i\hbar}[\hat{r}^{a}_{\bm{k}}, \hat{v}^{c}_{\bm{k}}] {C}_{n_2\bm{k}}\,.
\end{equation}
Note that the wave vector has only in-plane components $x$ and $y$, and the derivative $\frac{\partial}{\partial k^{z}}$ thus gives zero and $(r^{a}_{k})_{;n_1n_2k^{z}} = -i(\xi^{z}_{n_1n_1k} - \xi^{z}_{n_2n_2k})r^{a}_{n_1n_2k}$.

\subsection{SHG and EFISH Conductivities}
We focus on the SHG and EFISH induced by a uniform optical field $\bm{E}(t) = \bm{E}_0(t)e^{-i\omega t} + \text{c.c.}$, where $\bm{E}_0(t)$ is a slowly varying envelope function. By using the electronic states described above, the electron dynamics under applying an optical field can be determined by solving the semiconductor Bloch equation in the length gauge\, \cite{Cheng2015, Aversa1995, Hipolito2016, Pedersen2015, Hipolito2018}. Employing a perturbation theory with respect to the electric field, the second and third-order optical conductivities can be obtained, and the details are shown in Appendix \,\ref{appendix}. We focus on the nonlinear conductivities of $\sigma^{(2);dab}(\omega, \omega)$ for SHG, and $\sigma^{(3);dabc}(\omega, \omega,0)$ for EFISH.

Before the numerical calculation, we analyze the symmetry properties of the conductivity tensors. Without applying a static electric field, both the crystal structures of monolayer and bilayer BP belong to the $D_\text{2h}$ point group\,\cite{Jiang2016, Boyd2008}. For the EFISH conductivity tensor $\sigma^{(3);dabc}(\omega,\omega,0)$, there exist in total 21 nonzero components, among which 15 are independent and they are
\begin{enumerate}
\item[] $xxxx$, $xyyx$, $xzzx$, $yyxx=yxyx$, $zzxx=zxzx$,
\item[] $xxyy=xyxy$, $yyyy$, $yxxy$, $yzzy$, $zzyy=zyzy$,
\item[] $xxzz=xzxz$, $yyzz=yzyz$, $zxxz$, $zyyz$, $zzzz$,
\end{enumerate}
where the internal permutation symmetry $\sigma^{(3);dabc}(\omega, \omega, 0)=\sigma^{(3);dbac}(\omega, \omega, 0)$ has been applied.
Then the effective SHG conductivity is calculated as
\begin{align}
\sigma_{\text{eff}}^{(2);dab}(\omega,\omega) = 3 E_{\text{dc}}^c \sigma^{(3);dabc}(\omega, \omega, 0)\,,
\label{s3ands2}
\end{align}
and the corresponding effective second-order nonlinear susceptibility is given by
\begin{align}
\chi^{(2);dab}_\text{eff}(2\omega) = -\frac{3E_{\text{dc}}^c \sigma^{(3);dabc}(\omega, \omega, 0)}{2i\omega\epsilon_0 D}\,,
\label{eff}
\end{align}
with the vacuum dielectric constant $\epsilon_0$, and the effective thickness $D$.
When a perpendicular static electric field or a gate voltage is applied, the symmetry of the electronic states are reduced to $C_\text{2v}$ point group\,\cite{Xia2019}, which breaks the inversion symmetry, and the second-order SHG susceptibility can exist.
Accordingly, there exist in total seven nonzero components for $\sigma^{(2);dab}(\omega,\omega)$, among which 5 are independent; these nonzero components are $xxz=xzx$, $yyz=yzy$, $zxx$, $zyy$, and $zzz$. These nonzero components are consistent with the effective second-order conductivities induced by EFISH process by taking the static electric field is along the $z$ direction. 

\section{RESULTS FOR MONOLAYER BLACK PHOSPHORENE}
\label{mono}

\subsection{EFISH Response}

\begin{figure*}[!htp]
	\centering
	\includegraphics[scale=1]{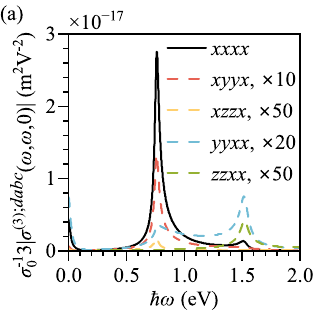}
	\includegraphics[scale=1]{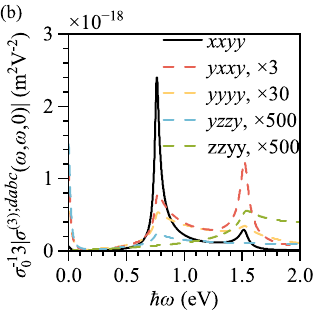}
	\includegraphics[scale=1]{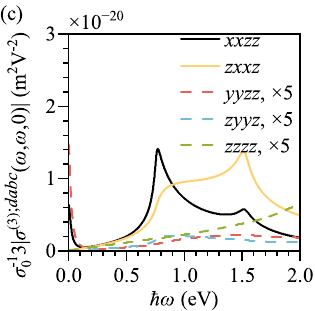}
	\caption{Spectra of $3\sigma_0^{-1}|\sigma^{(3);dabc}|$ for monolayer BP with the static electric field direction along the  (a) $x$-, (b) $y$-, and (c) $z$-directions, respectively. Here $\sigma_0 = \frac{e^2}{4 \hbar}$. Along each direction, the large elements are plotted with solid lines, specifically corresponding to $\sigma^{(3);xxxx}(\omega, \omega,0)$, $\sigma^{(3);xxyy}(\omega, \omega,0)$, $\sigma^{(3);xxzz}(\omega, \omega,0)$, and $\sigma^{(3);zxxz}(\omega, \omega,0)$; while other components are plotted by dashed lines scaled by a factor for visibility.}
	\label{mo}
\end{figure*}
In numerically calculating the EFISH or SHG conductivities, the Brillouin zone is divided into a uniform 1000$\times$1000 grid, which gives converged results. Other parameters are set as $T = 300$ K and $\gamma = 33$ {meV}. The monolayer BP has a bandgap of $\epsilon^\text{ML}_\text{g} = 1.52$ eV, located at the $\Gamma$ point, where the conduction band minimum is $\epsilon_{\text{c}\Gamma} = 0.34$ eV and the valence band maximum is $\epsilon_{\text{v}\Gamma} = -1.18$ eV.

Figure\,\ref{mo} presents the spectra of the all tensor components of EFISH conductivity $\sigma^{(3)}(\omega, \omega,0)$ for $\hbar\omega < 2.0 \, \text{eV}$ with a chemical potential $\mu = 0 $ {eV}. 
The spectra exhibit several features as follows:
(1) For $E_{\text{dc}} \parallel \hat{\bm x}$, the largest component among the induced conductivity $xxxx, xyyx, xzzx, yyxx = yxyx, zzxx = zxzx$ is $xxxx$, which is an order of magnitude larger than other components; thus the generated SH current is mainly along the $x$ direction (armchair direction).
For $E_{\text{dc}} \parallel \hat{\bm y}$, the largest component is $xxyy=xyxy$ and again the generated SH current is mainly along the $x$ direction; however, the maximal values of $xxyy$ component is about one order of magnitude smaller than that of $xxxx$. For $E_{\text{dc}} \parallel \hat{\bm z}$, both $xxzz=xzxz$ and $zxxz$ give relative large value; however, both are about two orders of magnitude smaller than that of $xxyy$.
Therefore, these results confirm the strong anisotropy in the nonlinear optical response in BP, where the in-plane armchair direction has stronger response than other directions, which can be attributed to the intrinsic asymmetry of the band structure,similar to the linear absorption of monolayer BP\,\cite{Le2019}.
(2) The spectra of all components exhibit  pronounced resonant peaks at $\hbar \omega$ $\approx$ $\epsilon^\text{ML}_\text{g}/2$ = 0.76 eV and $\epsilon^\text{ML}_\text{g}$ = 1.52 eV, corresponding to the two- and one-photon resonant transition at the bandgap.
(3) We estimate the effective SHG susceptibility using Eq.\,(\ref{eff}), with the thickness of the monolayer BP $D = c = 5.46$ {\AA}.
At $\hbar\omega = 0.76$ {eV}, the maximum of the conductivity gives $\sigma^{-1}_{0}3\left|\sigma^{(3);xxxx} \right|=2.75 \times 10^{-17}$ $\text{m}^{2}/\text{V}^{2}$, a static electric field $E_{\text{dc}}^c= 10^{7}$ {V/m} gives the effective SHG susceptibility as $\left|\chi^{(2)}_\text{eff}(2\omega) \right| = 1.5 \times 10^{3}$ {pm/V}, which is about 22 times larger than the widely used nonlinear crystal AgGaSe$_2$ (68 pm/V)\,\cite{Wu2012}.

\subsection{SHG at Different Gate Voltages}

\begin{figure*}[!htp]
	\centering
	\includegraphics[scale=1]{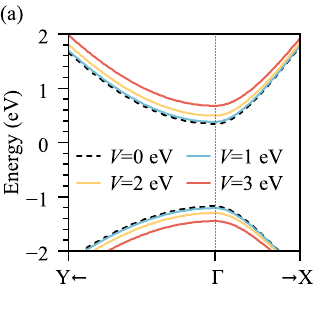}
	\includegraphics[scale=1]{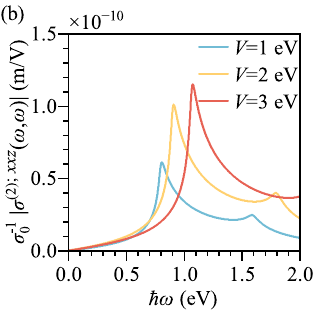}
	\includegraphics[scale=1]{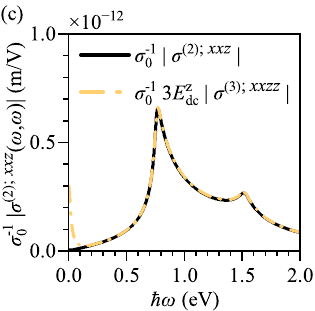}
	\caption{(a) The band structure of monolayer BP for different gate voltage V. (b) The spectra of the conductivity $\sigma^{(2);xxz}(\omega, \omega)$ for monolayer BP for gate voltage $V=1,2,3$ eV. (c) Comparison between $\sigma^{(2);xxz}(\omega, \omega)$ for a gate voltage 0.01 eV and $3E_{\text{dc}}^c\sigma^{(3);xxzz}(\omega, \omega,0)$ with $E_{\text{dc}}^c= 4.7\times10^{7}$ m/V.}
	\label{mov}
\end{figure*}

Now we reinvestigate the SHG under a gate voltage of up to 3 eV, which is feasible in experiments \cite{Pham2019}.
In this case, the effects of the static field are included in the band structure.
Figure\,\ref{mov}\,(a) shows the band structure for different gate voltages $V=0, 1, 2$ and $3$ eV. 
As the gate voltage increases, the conduction band moves to higher energy and the valence band moves to lower energy, leading to an increase of the bandgap.
Correspondingly, the bandgaps become 1.52, 1.59, 1.80, and 2.13 eV, respectively. 
Note that the monolayer BP maintains a direct bandgap under all applied voltages.
In Fig.\,\ref{mov}\,(c), we compare the spectra of $\sigma^{(2);xxz}$ for $V = 0.01$ eV, which corresponds to electric field $E_\text{dc}^z = 4.7 \times 10^7$ V/m, and $3 E_\text{dc}^z \sigma^{(3);xxzz}$ by Eq.\,(\ref{s3ands2}); and they are almost the same. 
The results indicate that the SHG and EFISH are consistent for very low gate voltage. 
For large gate voltage, the bandgap significant changes, and it means that the EFISH cannot be used to understand the SHG process. 
Figure\,\ref{mov}\,(b) shows the absolute values of the spectra of $\sigma^{(2);xxz}(\omega, \omega)$ under different gate voltages associated with Fig.\,\ref{mov}\,(a).
With the increase of gate voltage, the resonant peak location clearly blue shifts due to the increase in bandgap, and the peak intensity increases. 

\subsection{Chemical Potential Dependence of SHG and EFISH Conductivities}
\label{MONOLAYER-Chemical potential doping}

\begin{figure*}[htp]
	\centering
	\includegraphics[scale=1]{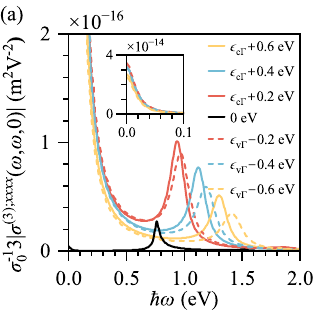}
	\includegraphics[scale=1]{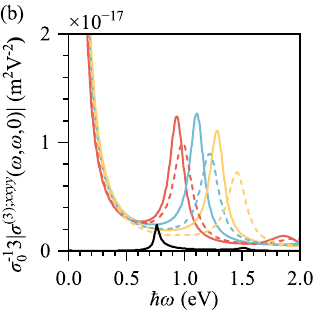}\\
	\includegraphics[scale=1]{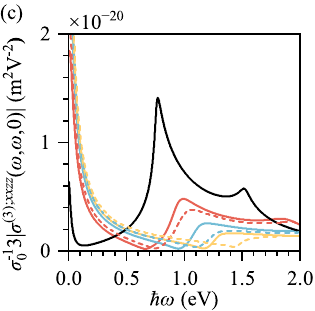}
	\includegraphics[scale=1]{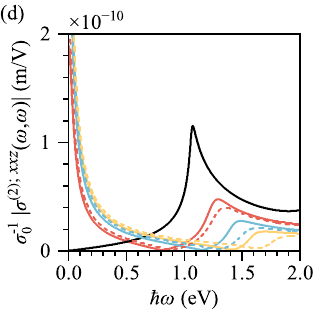}
	\caption{The spectra of (a) $\sigma^{(3);xxxx}(\omega, \omega,0)$, (b) $\sigma^{(3);xxyy}(\omega, \omega,0)$, (c) $\sigma^{(3);xxzz}(\omega, \omega,0)$, and (d) $\sigma^{(2);xxz}(\omega, \omega)$ at gate voltage of $V=3$ eV, for different doping levels  $\mu=\epsilon_{c\Gamma}+0.6$ eV, $\epsilon_{c\Gamma}+0.4$ eV, $\epsilon_{c\Gamma}+0.2$ eV, 0 eV, $\epsilon_{v\Gamma}-0.2$ eV, $\epsilon_{v\Gamma}-0.4$ eV, and $\epsilon_{v\Gamma}-0.6$ eV.}
	\label{momu}
\end{figure*}

Figure\,\ref{momu} gives the chemical potential denpendent spectra of $\sigma^{(3);xxxx}(\omega, \omega,0)$, $\sigma^{(3);xxyy}(\omega, \omega,0)$,  $\sigma^{(3);xxzz}(\omega, \omega,0)$, and $\sigma^{(2);xxz}(\omega, \omega)$ at gate voltage of $V=3$ eV, where we choose 
$\mu=\epsilon_{c\Gamma}+0.6$ eV, $\epsilon_{c\Gamma}+0.4$ eV, $\epsilon_{c\Gamma}+0.2$ eV, 0 eV, $\epsilon_{v\Gamma}-0.2$ eV, $\epsilon_{v\Gamma}-0.4$ eV, and $\epsilon_{v\Gamma}-0.6$ eV to keep the same doping levels for EFISH and SHG.
The chemical potential affects their conductivities in three aspects: 
(1) For nonzero $\mu$ describing the extra free electron/hole density, a significant resonant peak appears as the photon energy goes to zero, which arises from Drude-type divergences induced by the free carriers. 
For our parameter $\Gamma=33$ meV, $\sigma^{(3);xxxx}(\omega, \omega,0)$ can be as large as about $7\times10^{-19}$ S$\cdot$m$^2\cdot$V$^{-2}$.
(2) For intrinsic BP, there exists a peak around $\hbar \omega = \epsilon^\text{ML}_\text{g}/2$, which is induced by a two-photon resonant transition at the band edges. 
For all chemical potentials considered, optical transitions at the band edge are Pauli blocked. 
Nevertheless, a prominent peak still exists, which is attributed to the photon energy associated with interband transitions from the Fermi surface.
With the increase of the chemical potential, this peak location blue shifts.
(3) Compared with the undoping case, the nonzero $\mu$ leads to the enhancement of  $\sigma^{(3);xxxx}(\omega, \omega,0)$ and $\sigma^{(3);xxyy}(\omega, \omega,0)$, while the weakness of  $\sigma^{(3);xxzz}(\omega, \omega,0)$ and $\sigma^{(2);xxz}(\omega, \omega)$.

\section{RESULTS FOR BILAYER BLACK PHOSPHORENE}
\label{bl}

\subsection{EFISH Response}

\begin{figure*}[!htp]
	\centering
	\includegraphics[scale=1]{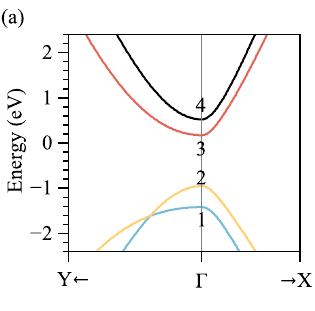}
	\includegraphics[scale=1]{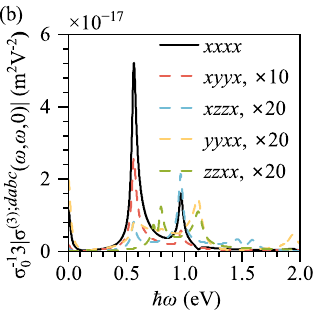}\\
	\includegraphics[scale=1]{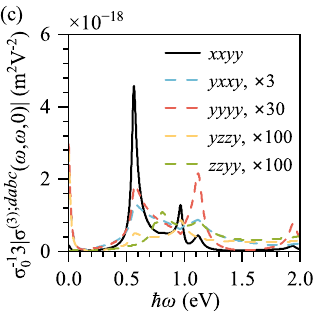}
	\includegraphics[scale=1]{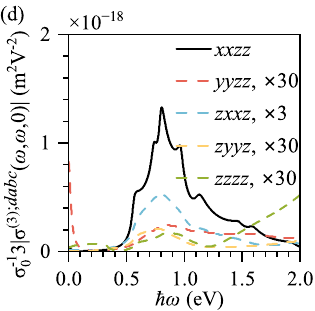}
	\caption{(a) Band structure of bilayer BP. (b)-(d) Spectra of absolute EFISH conductivity $|\sigma^{(3)}(\omega, \omega,0)|$ for bilayer BP with 15 tensor elements. The largest elements under each gate voltage are shown with black solid lines, while the smaller components are indicated by dashed lines.}
	\label{bilayer}
\end{figure*}
Now we turn to the EFISH of the bilayer BP. Figure\,\ref{bilayer}(a) illustrates the band structure of the bilayer BP near the band edge.
The bandgap is $\epsilon^\text{BL}_\text{g} = 1.12$ eV, which is less than that of the monolayer due to the interlayer coupling.
The EFISH spectra, analogous to those in Figs.\,\ref{bilayer}(b)-(d), show the conductivity spectra of EFISH for intrinsic bilayer BP, with the same parameter using in Figs.\,\ref{mo}(a)-(c).
It is easy to find that the spectra of $\sigma^{(3);xxxx}(\omega, \omega,0)$, $\sigma^{(3);xxyy}(\omega, \omega,0)$, and $\sigma^{(3);xxzz}(\omega, \omega,0)$ show much larger values than other components, which are very similar to the case in monolayer BP; and $\sigma^{(3);xxxx}(\omega, \omega,0)$ is the largest one. 
The spectra of $\sigma^{(3);xxxx}(\omega, \omega,0)$ and $\sigma^{(3);xxyy}(\omega, \omega,0)$ exhibit two distinct peaks, where one peak with larger amplitude locates at $\hbar \omega$  $\approx$ $\epsilon^\text{BL}_\text{g}/2$ = $(\epsilon_{3\Gamma}-\epsilon_{2\Gamma})/2$ = 0.56 {eV} associated with the interband two-photon resonant transition between the band edges of bands 2 and 3, and the other peak locates at $\hbar \omega$  $\approx$ $\epsilon^\text{BL}_\text{1-4}/2$ = $(\epsilon_{4\Gamma}-\epsilon_{1\Gamma})/2$ = 0.97 {eV} arising from the interband two-photon resonant transition between the band edges of bands 1 and 4. 
The $\sigma^{(3);xxxx}(\omega, \omega,0)$ and $\sigma^{(3);xxyy}(\omega, \omega,0)$ components also exhibit two smaller peaks located at $\hbar \omega \approx \epsilon^\text{BL}_\text{g} = \epsilon_{3\Gamma} - \epsilon_{2\Gamma} = 1.12$ eV and $\hbar \omega \approx \epsilon^\text{BL}_{1\text{-}4} = \epsilon_{4\Gamma} - \epsilon_{1\Gamma} = 1.94$ eV, respectively.
Different from $\sigma^{(3);xxxx}(\omega, \omega,0)$ and $\sigma^{(3);xxyy}(\omega, \omega,0)$, the spectra of $\sigma^{(3);xxzz}(\omega, \omega,0)$ show a complex multi-peak structure, which includes all possible one- and two-photon resonant transition between all bands. 
Compared with the results in the monolayer, obviously, the amplitudes of $\sigma^{(3);xxxx}(\omega, \omega,0)$ and $\sigma^{(3);xxyy}(\omega, \omega,0)$ in bilayer BP are nearly twice as large, while the $\sigma^{(3);xxzz}(\omega, \omega,0)$ is more than an order of magnitude higher, indicating the significant effects from interlayer coupling.

\subsection{SHG at Different Gate Voltages}

\begin{figure*}[!htp]
	\centering
	\includegraphics[scale=1]{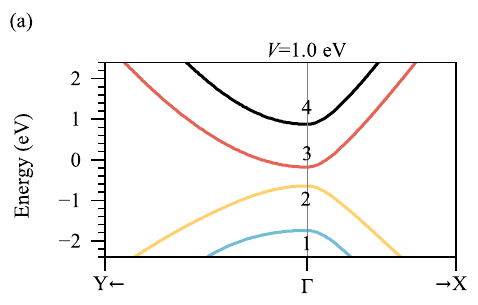}
	\includegraphics[scale=1]{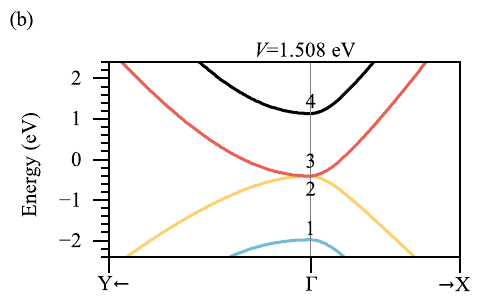}\\
	\includegraphics[scale=1]{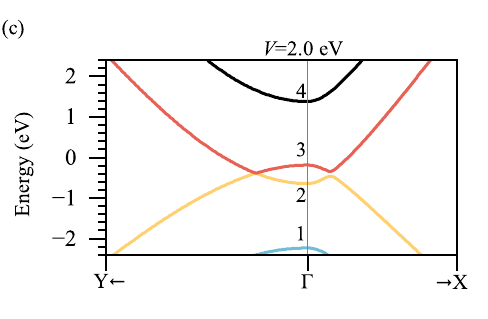}
	\includegraphics[scale=1]{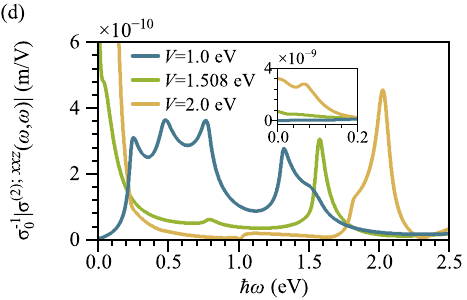}\\
	\includegraphics[scale=1]{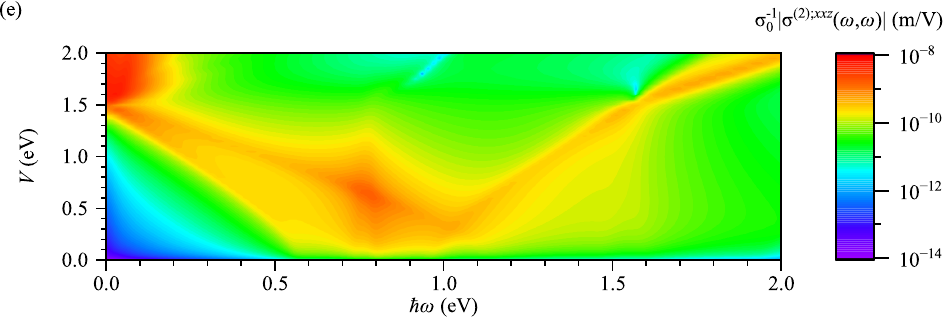}
	\caption{Band structures of bilayer BP under vertical gate voltages with (a) $V = 1.0$ eV, (b) $V = 1.508$ eV, and (c) $V = 2.0$ eV. (d) Spectra of the $\sigma^{(2);xxz}(\omega, \omega)$ component under three representative vertical gate voltages.
 (e) The gate voltage  dependent spectra of $\sigma^{(2);xxz}(\omega, \omega)$. } 
	\label{biv}
\end{figure*}

We then investigate the SHG conductivity $\sigma^{(2)}(\omega,\omega)$ for bilayer BP with applying different gate voltages.
Figure\,\ref{biv}\,(a)-(c) show band structures under three representative gate voltages $V$ = $1.0$, $1.508$, and $2.0$ eV, respectively.
As the gate voltage increases, the bands 1 and 3 move to lower energies, and the bands 2 and 4 move to higher energies.
For example, at $V = 1.0$ eV, as shown in Fig.\,\ref{biv}\,(a), the bandgaps are $\epsilon_{3\Gamma}-\epsilon_{2\Gamma}$ = 0.46 eV. 
At the critical gate voltage of $V_\text{c} = 1.508$ eV in Fig.\,\ref{biv}\,(b), the band edges of bands 2 and 3 coincide at the $\Gamma$ point, leading to the closure of bandgap, indicating a semiconducting-to-semimetal transition. 
When the gate voltage increases to $V = 2$ eV, as shown in Fig.\,\ref{biv}\,(c), the band touching point move to a $\bm k$ point along the path from $\Gamma$ to Y; the gap remains  zero and an inversion of the energy bands appears to form a Dirac cone.

Figure\,\ref{biv}\,(d) shows the spectra of $\sigma^{(2);xxz}(\omega, \omega)$ for the bilayer BP with the gate voltages mentioned above.
For $V=1$ eV, the bilayer BP is still semiconductor, and similar to the case of $\sigma^{(3);xxzz}(\omega, \omega, 0)$, there exist multiple resonant peaks, corresponding the one- and two-photon resonant transitions between different band edges. 
From lower to higher energies, the first two peaks appear at 0.23 eV and 0.46 eV, corresponding to two-photon and one-photon transitions, respectively, between the band edges of bands 2 and 3; the third peak, located at 0.76 eV, arises from two-photon resonant transitions involving both the band edges of bands 1 and 3, as well as bands 2 and 4; the fourth peak, observed at 1.31 eV, is induced by a two-photon resonant transition between bands 1 and 4.
With the application of a voltage that closes the bandgap (e.g., $V = 1.508$ eV or $V = 2.0$ eV), the first two resonant peaks disappear, while the third and fourth interband transition peaks undergo a blue shift.
Specifically,  as the voltage increases, the intensity of the third peak greatly decreases, while that of the fourth peak increases. 
Figure\,\ref{biv}\,(e) shows the spectra of the  $\sigma^{(2);xxz}(\omega, \omega)$ with more different gate voltages.
It is clear that 
the first two resonant peaks exhibit a red shift as the bandgap decreases, while the latter two peaks show a blue shift due to the increasing separation between the energy bands.
When the second and third peaks intersect (at $\hbar\omega \approx 0.8$ eV), the interband transition response reaches its maximum intensity.
Notably, when $1.5\ \text{eV} < V < 2\ \text{eV}$ (i.e., in bandgap-closure regime), a significant enhancement in the response is observed within the lower energy region ($\hbar\omega < 0.1$ eV).
This phenomenon arises from intraband transition resonant peaks induced by graphene-like band structures, whose intensity exceeds that of interband transition peaks by one order of magnitude.
Thus, due to the significant tuning effect of applying vertical gate voltage on bilayer band structure, the SHG conductivity exhibits different results.

\subsection{Chemical Potential Dependence of SHG and EFISH Conductivities}

\begin{figure*}[!htp]
	\centering
	\includegraphics[scale=1]{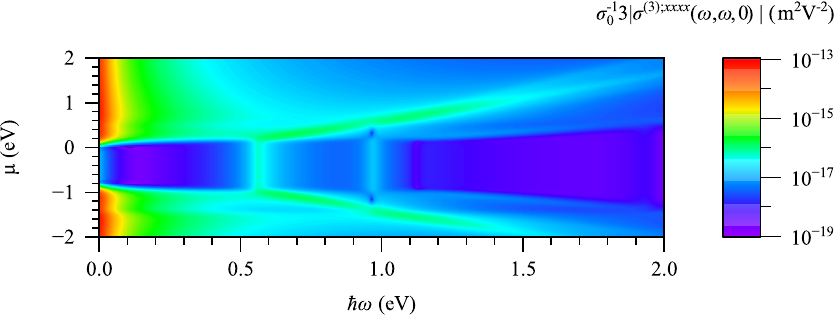}
	\caption{Chemical potential dependent spectra of $\sigma^{(3);xxxx}(\omega, \omega ,0)$. }
	\label{bimu-xxxx}
\end{figure*}

\begin{figure*}[!htp]
	\centering
	\includegraphics[scale=1]{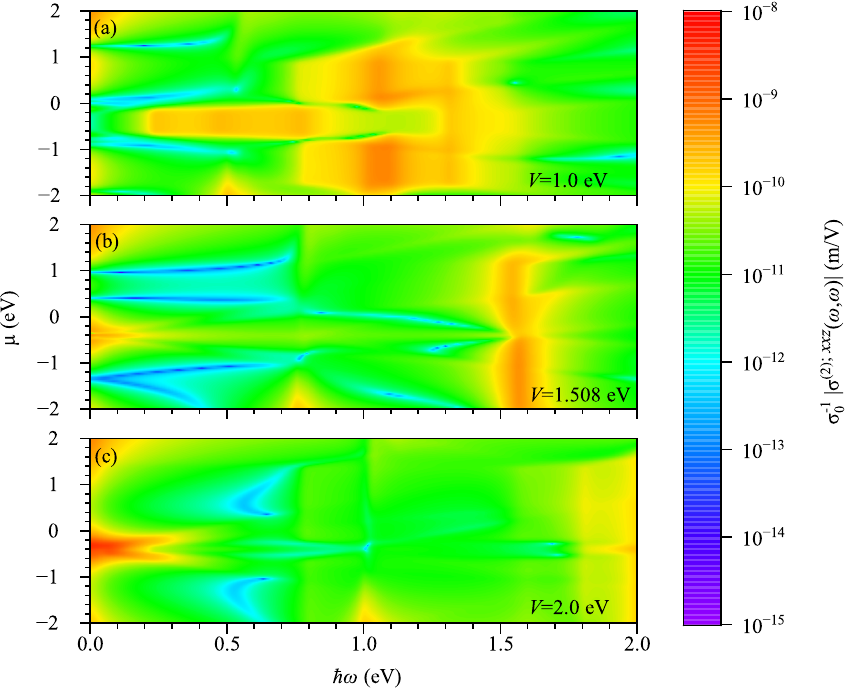}
	\caption{Chemical potential dependent spectra of $\sigma^{(2);xxz}(\omega, \omega)$ under three typical gate voltages with (a) $V = 1.0$ eV, (b) $V = 1.508$ eV and (c) $V=2.0$ eV. }
	\label{bimu}
\end{figure*}

Figure\,\ref{bimu-xxxx} shows $\sigma^{(3);xxxx}(\omega, \omega ,0)$ component of the bilayer BP without gate voltage.  
It reveals that the two bandgap resonant peaks exhibit a blue shift as the chemical potential moves upward or downward, yet remain unchanged when the chemical potential lies within the bandgap. This chemical potential tuning behavior is similar to that observed in the monolayer BP.

Figure\,\ref{bimu} shows the spectra of $\sigma^{(2);xxz}(\omega, \omega)$ as function of the chemical potential under the three aforementioned voltages.
In Fig.\,\ref{bimu}\,(a) with $V=1.0$ eV, it can be found that 
when  $\mu$ is within the bandgap ($\epsilon_{2\Gamma} < \mu < \epsilon_{3\Gamma}$),  the spectra basically remain  unchanged;
when  $\epsilon_{1\Gamma} < \mu < \epsilon_{2\Gamma}$  (or $\epsilon_{3\Gamma} < \mu < \epsilon_{4\Gamma}$), the new distinct resonant peaks referring to transitions within the valence (or conduction) bands appear in a range near $\hbar\omega = 1$ eV (here, ${\epsilon^{\text{BL}}_\text{1-2}}$ = 1.10 eV and ${\epsilon^{\text{BL}}_\text{3-4}}$ = 1.06 eV), and the transitions associated with the bandgaps vanish.
While, the chemical potential dependent changes at $V = 1.508$ eV and $V = 2.0$ eV are not significant, as shown in Figs.\,\ref{bimu}\,(b, c). 
The reason could be that the main peaks of their spectra are from the transitions at $\Gamma$ point, and $\epsilon_{2\Gamma}-\epsilon_{1\Gamma}\approx\epsilon_{4\Gamma}-\epsilon_{3\Gamma}\approx(\epsilon_{4\Gamma}-\epsilon_{1\Gamma})/2$. Therefore, 
the peaks with doping cannot be well distinguished from those without doping. 
Additionally, the transitions in the low-energy region occur within the range of $\mu$ from $-1$ eV to $0$ eV.

By comparing the influence of voltage and chemical potential on the $\sigma^{(3);xxxx}(\omega, \omega, 0)$ and $\sigma^{(2);xxz}(\omega, \omega)$ components in the bilayer BP, it can be observed that under a weak electric field, the interband transition resonant peaks of the $\sigma^{(3);xxxx}(\omega, \omega, 0)$ component can be effectively tuned by the chemical potential, as shown in Fig.\,\ref{bimu-xxxx}.
However, when a larger gate voltage is applied, the tuning of the interband resonant peaks in the $\sigma^{(2);xxz}(\omega, \omega)$ component depends mainly on the voltage-induced band structure modification, as shown in Fig.\,\ref{biv}\,(e). 
In contrast, changes in the chemical potential lead to not only the disappearance of the interband resonant peaks, but also the emergence of new interband transitions, as illustrated in Fig.\,\ref{bimu}\,(a). 
These results reveal that in the bilayer BP, the gate voltage and the chemical potential play distinct roles under different conditions, resulting in a more complex tuning behavior compared to the monolayer BP.

\section{CONCLUSIONS}
\label{conclusions}

This work  investigates the SHG of monolayer and bilayer BP under the influence of gate voltage and chemical potential.
The electronic states are described using a tight-binding model, and the dynamics is simulated via the semiconductor Bloch equations.
Our results show that BP exhibits dominant nonlinear conductivity along the armchair direction, fully reflecting its intrinsic structural and electronic anisotropy. Simultaneously, the resonant enhancement can be found when the incident photon energy approaches half the bandgap, which highlights the crucial role of two-photon transitions in the nonlinear process. Most importantly, we discuss an electric-field-induced topological phase transition in bilayer BP: as the gate voltage increases, the system undergoes a transition from semiconductor to semimetal, ultimately forming anisotropic Dirac cones accompanied by bandgap closure and band inversion. 
This phase transition significantly alters its nonlinear optical response -- after the transition, interband transition resonant peaks are replaced by strong SHG signals originating from intraband transitions of Dirac fermions near zero photon energy, with an intensity enhancement exceeding one order of magnitude. This indicates that bilayer BP can be utilized for electrically tunable nonlinear sources covering spectra ranges from the near-infrared to the terahertz regime. 
Furthermore, chemical potential enables precise regulation of nonlinear responses either independently or synergistically with gate voltage, providing additional degrees of freedom for functional device design. 
We also investigate the relation between EFISH and SHG across different gate voltages,
where under lower gate voltage, the two exhibit good agreement, whereas 
under higher gate voltage, SHG is more feasible mechanism.

In summary, this work not only establishes a theoretical foundation for BP-based tunable nonlinear photonic devices (such as electro-optic modulators, frequency converters, and terahertz sources), but also expands the research scope of topological nonlinear optics. 
It significantly promotes the understanding of nonlinear optical responses in anisotropic two-dimensional materials.
Future studies could further explore similar phenomena in strained structures, substrate-mediated effects, and multilayer configurations, with theoretical predictions to be validated through nonlinear spectroscopic experiments in the terahertz frequency range.

\begin{acknowledgments}
This work has been supported by National Natural Science Foundation of China Grant No. 12034003 (J.L.C.) and No. 21961132023 (L.W.).
\end{acknowledgments}

\appendix
\section{Expressions of nonlinear optical conductivities}
\label{appendix}

With the eigenenergies and the Berry connections,  the electron dynamics under an optical field is described by the semiconductor Bloch equations
\begin{align}
	i\hbar\frac{\partial \rho_{n_1n_2\bm k}(t)}{\partial t}
	= [\epsilon_{\bm k}, \rho_{\bm k}]_{n_1n_2}
	-eE^a [r^a_{\bm k},\rho_{\bm k}(t)]_{n_1n_2}
	-ieE^a(t)\left(\rho_{\bm k}(t)\right)_{;n_1n_2k_a}
	+i\hbar\left. \frac{\partial \rho_{n_1n_2\bm k}}{\partial t} \right|_{\text{scat}}\,.
\end{align}
The initial condition is taken as the thermal equilibrium state
\begin{align}
	\rho_{n_1n_2\bm k}(-\infty)
	=\delta_{n_1n_2}f_{n_1\bm k}\,, 
\end{align}
where $f_{n_1\bm k}=\frac{1}{1+e^{(\epsilon_{n_1\bm k}-\mu)/(k_BT)}}$ is the Fermi-Dirac distribution for a chemical potential $\mu$ and temperature $T$.
The density matrix is expanded as
\begin{align}
	\rho_{n_1n_2\bm k}(t) =\sum_{j=0}^\infty  \rho^{(j)}_{n_1n_2\bm k}(t)\,,
\end{align}
with $\rho^{(j)}_{n_1n_2\bm k}(t) \propto E^j$.
Then $\rho^{(j)}_{n_1n_2\bm k}(t)$ satisfies
\begin{align}
	i\hbar\frac{\partial \rho^{(j)}_{n_1n_2\bm k}}{\partial t}
	= [\epsilon_{\bm k}, \rho^{(j)}_{\bm k}]_{n_1n_2}
	-eE^a [r^a_{\bm k},\rho^{(j-1)}_{\bm k}]_{n_1n_2}
	-ieE^a(t)\big(\rho^{(j-1)}_{\bm k}\big)_{;n_1n_2k_a}
	+i\hbar\left. \frac{\partial \rho^{(j)}_{n_1n_2\bm k}}{\partial t} \right|_{\text{scat}}\,,
\end{align}
where $\rho^{(j)}_{\bm k} \equiv 0$ for $j < 0$. As a very rough approximation, a relaxation time approximation\,\cite{Cheng2015}  can be adopted to give
\begin{align}
	i\hbar\left. \frac{\partial \rho^{(j)}_{n_1n_2\bm k}}{\partial t} \right|_{\text{scat}}
	= -i\hbar\gamma_j\rho^{(j)}_{n_1n_2\bm k}
	,\quad\quad (j \geq 1)\,,
\end{align}
where $\gamma_j > 0$ is a relaxation parameter introduced to describe
the dynamics of $\rho^{(j)}_{n_1n_2\bm k}$.	
Explicitly we can write the $\rho ^{(n)}_{n_1n_2\bm k}(t)$ up to the third-order in terms of the electric field as
\begin{align}
	\rho ^{(1)}_{n_1n_2\bm k}(t)=
	\left(\frac{-e}{\hbar}\right)\int \frac{d\omega_3}{2\pi} 
	\mathcal{P}^{(1);c}_{n_1n_2\bm k}(w_3) E^c (\omega_3)e^{-i\omega_3 t}	\,,
\end{align}
\begin{align}
	\rho ^{(2)}_{n_1n_2\bm k}(t)
	=&\left(\frac{-e}{\hbar}\right)^2\int \frac{d\omega_2 d\omega_3}{(2\pi)^2}
	\mathcal{P}^{(2);bc}_{n_1n_2\bm k}(w_0,w_3)
	E^b(\omega_2)E^c(\omega_3)
	e^{-i(\omega_2+\omega_3) t}\,,
\end{align}
\begin{align}
	\rho ^{(3)}_{n_1n_2\bm k}(t)
	=&\left(\frac{-e}{\hbar}\right)^3\int \frac{d\omega_1 d\omega_2 d\omega_3}{(2\pi)^3}
	\mathcal{P}^{(3);abc}_{n_1n_2\bm k}(w,w_0,w_3)
	E^a(\omega_1)E^b(\omega_2)E^c(\omega_3)
	e^{-i(\omega_1+\omega_2+\omega_3) t}\,.
\end{align}
Here,
\begin{align}
	\mathcal{P}^{(1);c}_{n_1n_2\bm k}(w_3)=
	\frac{[r^c_{\bm k},f_{\bm k}]_{n_1n_2}+i\big(f_{\bm k}\big)_{;n_1n_2k_c}}{w_3-\omega_{n_1n_2\bm k}}\,,
\end{align}
\begin{align}
	\mathcal{P}^{(2);bc}_{n_1n_2\bm k}(w_0,w_3)=
	\frac{[r^b_{\bm k},\mathcal{P}^{(1);c}_{\bm k}(w_3)]_{n_1n_2}+i{(\mathcal{P}^{(1);c}_{\bm k}(w_3))}_{;n_1n_2k_b}}{w_0-\omega_{n_1n_2\bm k}}\,,
\end{align}
\begin{align}
	\mathcal{P}^{(3);abc}_{n_1n_2\bm k}(w,w_0,w_3)=
	\frac{[r^a_{\bm k},\mathcal{P}^{(2);bc}_{\bm k}(w_0,w_3)]_{n_1n_2}+i{(\mathcal{P}^{(2);bc}_{\bm k}(w_0,w_3))}_{;n_1n_2k_a}}{w-\omega_{n_1n_2\bm k}} \,.
\end{align}
We define $w_3=\omega_3+i\gamma_1$, $w_0=\omega_2+\omega_3+i\gamma_2$, and $w=\omega_1+\omega_2+\omega_3+i\gamma_3$, and assume $\gamma_1 = \gamma_2 = \gamma_3 = \gamma$. The covariant derivative is expressed as
$(X_k)_{;n_1n_2k_a} = \frac{\partial X_{n_1n_2k}}{\partial k_a} - i(\xi^n_{n_1n_1k} - \xi^n_{n_2n_2k})X_{n_1n_2k}$.

The expansion of current density is
\begin{align}
	\bm{J}=\bm{J}^{(1)}(t)+\bm{J}^{(2)}(t)+\bm{J}^{(3)}(t)+\cdots,
\end{align}
with
\begin{align}
	J^{(n);d}(t)=e\sum_{n_1n_2}\int \frac{d\bm k}{(2\pi)^2} v^d_{n_2n_1\bm k} \rho ^{(n)}_{n_1n_2\bm k}(t).
\end{align}
We are interested in the current density up to the third-order, defined as
\begin{align}
	J^{(1);d} = 
	\int \frac{d\omega_3}{2\pi}
	{\sigma}^{(1);dc}(\omega_3)
	E^c (\omega_3)e^{-i\omega_3 t}\,,
\end{align}
\begin{align}
	J^{(2);d} = 
	\int \frac{d\omega_2d\omega_3}{(2\pi)^2}
	{\sigma}^{(2);dbc}(\omega_2,\omega_3)
	E^c (\omega_3)E^b (\omega_2)e^{-i(\omega_3+\omega_2 )t}\,,
\end{align}
\begin{align}
	J^{(3);d} = 
	\int \frac{d\omega_1d\omega_2d\omega_3}{(2\pi)^3}
	{\sigma}^{(3);dabc}(\omega_1,\omega_2,\omega_3)
	&E^a(\omega_1)E^b(\omega_2)E^c(\omega_3)e^{-i(\omega_1+\omega_2+\omega_3)t}\,,
\end{align}
with 
\begin{equation}
	\sigma^{(2);dab}(\omega_1, \omega_2) = \frac{1}{2} \left[ \tilde{\sigma}^{(2);dab}(\omega_0 + i\gamma_2, \omega_2 + i\gamma_1) + \tilde{\sigma}^{(2);dba}(\omega_0 + i\gamma_2, \omega_1 + i\gamma_1) \right]\,,
\end{equation}
and
\begin{equation}
	\begin{aligned}
		\sigma^{(3);dabc}(\omega_1, \omega_2, \omega_3) = \frac{1}{6} \Bigl[ 
		& \tilde{\sigma}^{(3);dabc}(\omega + i\gamma_3, \omega_2 + \omega_3 + i\gamma_2, \omega_3 + i\gamma_1) \\
		& + \tilde{\sigma}^{(3);dacb}(\omega + i\gamma_3, \omega_2 + \omega_3 + i\gamma_2, \omega_2 + i\gamma_1) \\
		& + \tilde{\sigma}^{(3);dbac}(\omega + i\gamma_3, \omega_1 + \omega_3 + i\gamma_2, \omega_3 + i\gamma_1) \\
		& + \tilde{\sigma}^{(3);dbca}(\omega + i\gamma_3, \omega_1 + \omega_3 + i\gamma_2, \omega_1 + i\gamma_1) \\
		& + \tilde{\sigma}^{(3);dcab}(\omega + i\gamma_3, \omega_1 + \omega_2 + i\gamma_2, \omega_2 + i\gamma_1) \\
		& + \tilde{\sigma}^{(3);dcba}(\omega + i\gamma_3, \omega_1 + \omega_2 + i\gamma_2, \omega_1 + i\gamma_1) \Bigr]\,.
	\end{aligned}
\end{equation}
Here $\tilde{\sigma}^{(2);dbc}(w_0,w_3)$ and $\tilde{\sigma}^{(3);dabc}(w,w_0,w_3)$ are unsymmetrized nonlinear conductivities
\begin{align}
	\tilde{\sigma}^{(1);dc}(w_3)
	=
	\left(\frac{-e^2}{\hbar}\right)
	\sum_{n_1n_2}\int \frac{d\bm k}{(2\pi)^2} 
	v^d_{n_2n_1\bm k} 
	\mathcal{P}^{(1);c}_{n_1n_2\bm k}(w_3)\,,
\end{align}
\begin{align}
	\tilde{\sigma}^{(2);dbc}(w_0,w_3)
	=
	\left(\frac{e^3}{{\hbar}^2}\right)
	\sum_{n_1n_2}\int \frac{d\bm k}{(2\pi)^2} 
	v^d_{n_2n_1\bm k} 
	\mathcal{P}^{(2);bc}_{n_1n_2\bm k}(w_0,w_3)\,,	
\end{align}
\begin{align}
	\tilde{\sigma}^{(3);dabc}(w,w_0,w_3)
	=
	\left(\frac{-e^4}{{\hbar}^3}\right)
	\sum_{n_1n_2}\int \frac{d\bm k}{(2\pi)^2} 
	v^d_{n_2n_1\bm k} 
	\mathcal{P}^{(3);abc}_{n_1n_2\bm k}(w,w_0,w_3)\,.
\end{align}
The velocity matrix elements and Berry connections are calculated by Eq.\,(\ref{v}) and Eq.\,(\ref{r}), respectively.
The physical phenomenological relaxation parameters have to satisfy $\gamma_i > 0$, and the relaxation-free limit is obtained by setting $\gamma_i = 0^{+}$.
When taking $\omega_1 = \omega_2 = \omega$, $\sigma^{(2);dab}(\omega,\omega)$ represents the SHG. When taking $\omega_1 = \omega_2 = \omega$ and $\omega_3 = 0$, $\sigma^{(3);dabc}(\omega,\omega,0)$ represents EFISH.

\bibliography{BPref}
\end{document}